\documentclass[a4paper,12pt]{article}
\date {}

\usepackage{amsfonts}
\usepackage{amsmath}

\newcommand{\be}{\begin{eqnarray}}
\newcommand{\ee}{\end{eqnarray}}

\def\Ro{{\mathbb R}}
\def\qq{{\bf q}}
\def\pp{{\bf p}}

{}
{}
{}
{}
{}

\begin{document}

\title{Non-extensivity of the configurational density distribution
in the classical microcanonical ensemble}
\author{
Jan Naudts and Maarten Baeten\\
\small Departement Natuurkunde, Universiteit Antwerpen,\\
\small Universiteitsplein 1, 2610 Antwerpen, Belgium\\
\small E-mail Jan.Naudts@ua.ac.be, Maarten.Baeten@student.ua.ac.be
}
\maketitle
\begin{abstract}
We show that the configurational probability distribution of a classical gas
always belongs to the $q$-exponential family. Hence,
the configurational subsystem is non-extensive in the sense of Tsallis.
One of the consequences of this observation is that
the thermodynamics of the configurational subsystem is uniquely determined
up to a scaling function. As an example we consider a system of non-interacting
harmonic oscillators. In this example, the scaling function can be determined
from the requirement that in the limit of large systems the microcanonical temperature of the
configurational subsystem should coincide with that of the canonical ensemble.
The result suggests that R\'enyi's entropy function is the
relevant one rather than that of Tsallis.

\end{abstract}

\section{Introduction}

Many-particle systems are usually studied in the canonical or grandcanonical ensemble.
But for small systems the equivalence of ensembles breaks down and it becomes
interesting to study the microcanonical ensemble. One of the intriguing questions
in this context is whether phase transitions can occur in the way proposed by Dieter Gross
in \cite {GD01}. We will not consider this question here but use it as a motivation
to study the thermodynamics of closed systems.
The arguments of Gross are a mixture of statistical physics applied
on microscopic models and of macroscopic thermodynamics. A more clean approach
requires that the thermodynamic formalism is derived from the statistical theory
rather than being used as a generally valid but unexplained addition.

The present paper exploits the characteristic of classical systems that the coordinates
and the conjugated momenta can be considered as two mutually interacting subsystems.
By considering a subsystem one avoids the difficulty that the probability
distribution of the microcanonical ensemble is a singular measure (a Dirac delta function).
By integrating out the momenta, which is always possible for a Hamiltonian which is
quadratic in the momenta, the configurational probability distribution results.
To our surprise, this distribution function is nonextensive in the sense
of Tsallis \cite {TC04}.
Apparently, nobody has made this observation
in the past twenty years. The implications are important. It places
nonextensive thermostatistics at the heart of statistical physics. In addition, many properties
of the configurational subsystem can now be studied in a systematic way.

The present paper makes use of the knowledge that the probability distributions
of nonextensive thermostatistics automatically induce the standard thermodynamical
formalism. This means in particular that the thermodynamic
entropy and the temperature of the configurational subsystem are fixed in a unique manner
up to a monotonic function, which is referred to as the scaling function.

In the next Section we discuss various choices of the microcanonical entropy function.
Section 3 introduces the notion of a $q$-exponential family of probability distributions.
In Section 4 the configurational probability distribution is calculated. It is shown
to belong to the $q$-exponential family. Section 5 deals with thermodynamic
relations. Section 6 treats a simple example and discusses the scaling function.
Finally, a short discussion follows in Section 7.

\section {Microcanonical entropies}

The entropy $S(U)$ which is most often used in the classical microcanonical
ensemble is
\be
S(U)=k_B\ln\omega(U),
\label {intro:boltzmann}
\ee
where $\omega(U)$ is the $N$-particle density of states. The latter is given by
\be
\omega(U)=\frac 1{h^{3N}}\int_{\Ro^{3N}}{\rm d}\pp_1\cdots{\rm d}\pp_N
\int_{\Ro^{3N}}{\rm d}\qq_1\cdots{\rm d}\qq_N
\delta(U-H(\qq,\pp)).
\ee
Here, $\qq_j$ is the position of the $j$-th particle and $\pp_j$ is the conjugated momentum,
$H(\qq,\pp)$ is the Hamiltonian. The constant $h$ is introduced for dimensional reasons.
This definition goes back to Boltzmann's idea of equal probability
of the microcanonical states and the corresponding well-known formula
$S=k_B\ln W$, where $W$ is the number of microcanonical states.
However, this choice of definition of entropy has some drawbacks.
For instance, for the pendulum the entropy $S(U)$ as a function of
internal energy $U$ is a piecewise convex function
instead of a concave function \cite {NJ05}.
The lack of concavity can be interpreted as a microcanonical instability \cite {GD01,GD90}.
But there is no physical reason why the pendulum should be classified as being instable
at all energies.

The shortcomings of Boltzmann's entropy have been noticed long ago.
A slightly different definition of entropy is \cite {SA48,PHT85}
(see also in \cite {SBJ06} the reference to the work of A. Schl\"uter )
\be
S(U)=k_B\ln\Omega(U),
\label {intro:pearson}
\ee
where $\Omega(U)$ is the integral of $\omega(U)$ and is given by
\be
\Omega(U)=\frac 1{h^{3N}}\int_{\Ro^{3N}}{\rm d}\pp_1\cdots{\rm d}\pp_N
\int_{\Ro^{3N}}{\rm d}\qq_1\cdots{\rm d}\qq_N
\Theta(U-H(\qq,\pp)).
\ee
Here, $\Theta(x)$ is Heaviside's function.
An immediate advantage of (\ref {intro:pearson}) is that
the resulting expression for the temperature $T$, 
defined by the thermodynamical formula
\be
\frac 1{T}=\frac {{\rm d}S}{{\rm d}U},
\label {intro:Tdef}
\ee
coincides with the notion of temperature as used by experimentalists.
Indeed, one finds
\be
k_BT=\frac {\Omega(U)}{\omega(U)}.
\label {intro:Tres}
\ee
For a harmonic oscillator the density of states $\omega(U)$ is a constant.
Hence, (\ref {intro:Tres}) implies $k_BT=U$, as wanted.
It is well-known
that for classical monoatomic gases the r.h.s.~of (\ref {intro:Tres}) coincides
with twice the average kinetic energy per degree of freedom.
This result is also derived below --- see (\ref {therm:avkinenrg}).
Its significance is that the
equipartition theorem, assigning $k_BT/2$ to each degree of
freedom, does hold for the kinetic energy also in the microcanonical ensemble.
Quite often the average kinetic energy per degree of freedom
is experimentally accessible and provides a unique way to measure
accurately the temperature of the system.

But also (\ref {intro:pearson}) and (\ref {intro:Tres}) are subject to criticism.
In small systems finite size corrections appear \cite {SBJ06,USC08} for a number
of reasons. As argued in \cite {USC08}, the problem is not the equipartition of the
kinetic energy over the various degrees of freedom, but the relation between
temperature and kinetic energy.

\section {Generalized exponential family}

Recently, the notion of a generalized exponential family has been
introduced both in the physics \cite {NJ05a,NJ06,OW08,NJ08b} and
in the mathematics \cite {NJ04,GD04,ES04,NJ08} literature.
It is shown in the next Section that the
configurational probability distributions of a classical real gas
in the microcanonical ensemble always belong to the
$q$-exponential family, which is a special case of the generalized exponential family.
A first observation in this direction was made in \cite {NJ08b}.

Fix a number $q$. The probability distribution $f_\theta(x)$
with parameter $\theta$ is said to belong to
the $q$-exponential family if it can be written as
\be
f_\theta(x)=c(x)\exp_q(-\alpha(\theta)-\theta H(x)),
\label {gef:def}
\ee
where the $q$-deformed logarithm \cite {TC94,NJ02} is defined by
\be
\exp_q(u)=[1+(1-q)u]_+^{1/(1-q)}.
\ee
The notation $[u]_+=\max\{0,u\}$ is used. In (\ref {gef:def})
it is important that $H(x)$ and $c(x)$ do not depend on the parameter $\theta$
and that the normalization constant $\alpha(\theta)$ does not depend on $x$.
In the limit $q=1$ the $q$-exponential function reduces to the natural exponential function.
The notion of the $q$-exponential family then reduces to the standard
notion of an exponential family.

Distributions belonging to the $q$-exponential family share a number
of properties which make it attractive to work with these distributions.
See for instance \cite {NJ08b}.
These properties are well-known to physicists because the
Boltzmann-Gibbs distribution
\be
f_\beta(x)=\frac 1Z\exp(-\beta H(x))
\ee
belongs to the standard exponential family, which corresponds with the choice $q=1$.
In particular, when $H(x)$ is the energy of a mechanical system,
and $f_\theta(x)$ belongs to the $q$-exponential family,
then there is a unique way to fit the statistical model into the context
of thermodynamics.

\section{The configurational probability distribut\-ion}

A classical model of $N$ particles is determined by the Hamiltonian
\be
H(\qq,\pp)\equiv H(\qq_1,\qq_2,\cdots \qq_N,\pp_1,\pp_2,\cdots \pp_N),
\ee
where $\qq_j$ is the position of the $j$-th particle and $\pp_j$ is the conjugated momentum.
The microcanonical ensemble is then described by the singular probability density function
\be
f_U(\qq,\pp)=\frac 1{\omega(U)}\delta(U-H(\qq,\pp)),
\ee
where $\delta(\cdot)$ is Dirac's delta function.
The normalization is so that
\be
1=\frac 1{h^{3N}}\int_{\Ro^{3N}}{\rm d}\pp_1\cdots{\rm d}\pp_N
\int_{\Ro^{3N}}{\rm d}\qq_1\cdots{\rm d}\qq_N
f_U(\qq,\pp).
\ee
The particles are enclosed in a box with volume $V$.
For simplicity, we take only one conserved quantity into account, namely the total energy.
Its value is fixed to $U$.

In the simplest case the Hamiltonian is of the form
\be
H(\qq,\pp)=
\frac 1{2m}\sum_{j=1}^N|p_j|^2+{\cal V}(\qq),
\ee
where ${\cal V}(\qq)$ is the potential energy due to interaction
among the particles and between the particles and the walls of the system.
It is then possible to integrate out the momenta.
This leads to the configurational probability distribution, which is given by
\be
f_U^{\rm conf}(\qq)
&=&\frac {1}{h^{3N}}\int_{\Ro^{3N}}{\rm d}\pp_1\cdots{\rm d}\pp_N\,f_U(\qq,\pp).
\ee
The normalization is so that
\be
1&=&\int_{\Ro^{3N}}{\rm d}\qq_1\cdots{\rm d}\qq_N
f_U^{\rm conf}(\qq).
\ee
Let $B(N)$ denote the volume of a sphere of radius 1 in dimension $N$.
A short calculation gives
\be
f_U^{\rm conf}(\qq)
&=&\frac {1}{h^{3N}}\frac 1{\omega(U)}\int_{\Ro^{3N}}{\rm d}\pp_1\cdots{\rm d}\pp_N\,\delta(U-H(\qq,\pp))\crcr
&=&\frac {1}{h^{3N}}\frac 1{\omega(U)}\frac {{\rm d}\,}{{\rm d}U}
\int_{\Ro^{3N}}{\rm d}\pp_1\cdots{\rm d}\pp_N\,\Theta\left(U-{\cal V}(\qq)-\frac 1{2m}\sum_{j=1}^N|p_j|^2\right)\crcr
&=&\frac {1}{h^{3N}}\frac 1{\omega(U)}(2m)^{3N/2}B(3N)
\frac {{\rm d}\,}{{\rm d}U}
[U-{\cal V}(\qq)]_+^{3N/2}\crcr
&=&\frac {1}{2h^{3N}}\frac {3N}{\omega(U)}(2m)^{3N/2}B(3N)
\left[U-{\cal V}(\qq)\right]_+^{\frac 32N-1}\crcr
&=&c_N\exp_q\left(-\alpha(\theta)-\theta {\cal V}(\qq)\right),
\label {cpd:pdf}
\ee
with
\be
c_N&=&\left(\frac {2m}{h^2}\right)^{3N/2},\cr
\theta&=&\frac 1{1-q}\frac 1{[\Gamma(3N/2)\omega(U)]^{1-q}},\crcr
\alpha(\theta)&=&\frac 32N-1-\theta U,\crcr
q&=&1-\frac 2{3N-2}.
\label {cpd:params}
\ee
For convenience we assume here
that $\omega(U)$ is a strictly increasing function of $U$ so that it can be
inverted to obtain $U$ as a function of $\theta$.
One concludes from (\ref {cpd:pdf}) that the configurational density function $f_U^{\rm conf}(\qq)$
of a classical gas in the microcanonical ensemble with parameter $U$ always belongs to the
$q$-exponential family with the constant $q$ given by (\ref {cpd:params}).

\section{Dual identities}

It is well-known that
the $q$-exponential distribution optimizes the Tsallis entropy \cite {TC88}
and that together with the configurational energy $U^{\rm conf}$ it
satisfies the thermodynamic duality relations \cite {NJ05a}.
As shown below, these identities imply the statement that
the ratio $\Omega(U)/\omega(U)$ equals the average kinetic energy.

Of course, also any monotonically increasing function of the Tsallis entropy will be optimized
by the same probability distributions. In particular, R\'enyi's alpha-entropy \cite {RA65,RA76},
given by
\be
I_\alpha(f)=\frac 1{1-\alpha}\ln c_N\int_{\Ro^{3N}}{\rm d}\qq_1\cdots{\rm d}\qq_N\,
\left(\frac {f(\qq)}{c_N}\right)^\alpha,
\ee
is such an equivalent entropy function.
For that reason we will write the configurational entropy $S^{\rm conf}$
as an unknown monotonic function of a quantity $\tilde S^{\rm conf}$, where the later is obtained
by maximizing the entropy functional $I(f)$.

An appropriate way of writing $I(f)$ is \cite {NJ08b} (assume $k_B=1$ for convenience)
\be
I(f)=-c_N\int_{\Ro^{3N}}{\rm d}\qq_1\cdots{\rm d}\qq_N\,
F\left(\frac 1{c_N}f(\qq)\right)
\label {dual:entropydef}
\ee
with
\be
F(u)=\int_0^u{\rm d}v\,\ln_q(v)=\frac u{1-q}\left(\frac 1{2-q}u^{1-q}-1\right).
\ee
Using (\ref {cpd:pdf}, \ref {cpd:params}) one obtains
\be
\tilde S^{\rm conf}&\equiv&
I(f_U^{\rm conf})\crcr
&=&-\frac 1{1-q}\int_{\Ro^{3N}}{\rm d}\qq_1\cdots{\rm d}\qq_N\,
f_U^{\rm conf}(\qq)
\left(\frac 1{2-q}\left(\frac 1{c_N}f_U^{\rm conf}(\qq)\right)^{1-q}-1\right)\crcr
&=&\frac 1{2-q}\left(1+\alpha(\theta)+\theta U^{\rm conf}\right)\crcr
&=&\frac 1{1-q}-\frac \theta{2-q} U^{\rm kin},
\label {therm:entropy}
\ee
with $U^{\rm kin}=U-U^{\rm conf}$.

The corresponding Massieu function is then given by
\be
\tilde \Phi(\theta)&=&\tilde S^{\rm conf}-\theta U^{\rm conf}\crcr
&=&\frac 1{1-q}-\frac {\theta U}{2-q}
-\frac {1-q}{2-q}\theta U^{\rm conf}.
\ee
Using the dual identities \cite {NJ05a,NJ08b}
\be
\frac {{\rm d}\tilde\Phi}{{\rm d}\theta}=-U^{\rm conf}
\quad\mbox{ and }\quad
\frac {{\rm d}\tilde S^{\rm conf}}{{\rm d}U^{\rm conf}}=\theta
\label {therm:dualrel}
\ee
one obtains
\be
U^{\rm conf}=U+(1-q)\theta \frac {{\rm d}U^{\rm conf}}{{\rm d}\theta}
+\theta\frac{{\rm d}U}{{\rm d}\theta}
\label {therm:dual1}
\ee
and
\be
(1-q)\theta
&=&-\left[
U^{\rm kin}\frac {{\rm d}\theta}{{\rm d}U}
+\theta
\right]\frac {{\rm d}U}{{\rm d}U^{\rm conf}}.
\label {therm:dual2}
\ee
Since both identities imply the same result we continue with one of them.
The latter can be written as
\be
\frac {{\rm d}U^{\rm conf}}{{\rm d}U}
=\frac {\omega'(U)}{\omega(U)}U^{\rm kin}-\frac 1{1-q}.
\ee
Use this result to calculate
\be
\frac {{\rm d}\,}{{\rm d}U}\omega(U)
U^{\rm kin}
&=&\omega'(U)U^{\rm kin}+\omega(U)\left[1
-\frac {{\rm d}U^{\rm conf}}{{\rm d}U}\right]\crcr
&=&\frac {2-q}{1-q}\omega(U)=\frac {3N}2\omega(U).
\ee
By integrating this expression one obtains
the average kinetic energy
\be
U^{\rm kin}
=\frac {3N}2\frac {\Omega(U)}{\omega(U)}.
\label {therm:avkinenrg}
\ee
This expression gives the relation between the average kinetic
energy and the total energy $U$ via the density of states $\omega(U)$ and its integral $\Omega(U)$.
The integration constant must be taken so that $\Omega(U)=0$ when $U=U_{\rm min}$
(implying that the kinetic energy vanishes in the ground state).

\section{Example}

Consider a set of $3N$ harmonic oscillators. The potential energy equals
\be
{\cal V}(\qq)=\frac12m \sum_{j=1}^{3N}\omega_j^2q_j^2.
\ee
One calculates
\be
\Omega(U)&=&\frac 1{h^{3N}}\int_{\Ro^{3N}}{\rm d}p_1\cdots{\rm d}p_{3N}
\int_{\Ro^{3N}}{\rm d}q_1\cdots{\rm d}q_{3N}\crcr
& &\times
\Theta\left(U-\frac 1{2m}\sum_jp_j^2-\frac 12m\sum_j\omega_j^2q_j^2
\right)\crcr
&=&\frac 1{(3N)!\prod_{j=1}^{3N}\omega_j}\left(\frac {4\pi U}{h}\right)^{3N}.
\label {ex:Omega}
\ee
From (\ref {cpd:params}) and (\ref {ex:Omega}) now follows
\be
\frac 1\theta=(1-q)\left[
\frac {\Gamma(3N/2)}{\Gamma(3N)}\frac 1{\prod_{j=1}^{3N}\omega_j}\left(\frac {4\pi }{h}\right)^{3N}
U^{3N-1}
\right]^{1-q}.
\ee
Using Stirling's approximation one obtains
\be
\frac {\Gamma(3N/2)}{\Gamma(3N)}
\sim\sqrt{2}\left(\frac {\rm e}{6N}\right)^{3N/2}
\ee
so that, assuming that all $\omega_j$ are equal to some $\omega$,
\be
\frac 1\theta\sim {\rm e}\left(\frac {4\pi}{h\omega}\frac U{3N}\right)^2.
\label {ho:restemp}
\ee
On the other hand, the inverse temperature $\beta$ of the canonical
ensemble with $3N$ degrees of freedom satisfies
\be
\frac 1\beta=\frac U{3N}.
\ee
The requirement that the canonical temperature coincides with the microcanonical
temperature as obtained from (\ref {intro:Tdef}) now determines the monotonic
function which relates the thermodynamic $S^{\rm conf}$ to the quantity $\tilde S^{\rm conf}$.
From this requirement follows that
\be
S^{\rm conf}=N\ln U^{\rm conf}+A,
\ee
for some constant $A$. On the other hand, (\ref {ho:restemp}) implies that
\be
\tilde S^{\rm conf}\sim -\frac 1{{\rm e}}\left(\frac {h\omega}{4\pi}\right)^2\frac {(3N)^2}{U^{\rm conf}}+
\mbox{constant.}
\label {ex:sappr}
\ee
Therefore the relation between $S^{\rm conf}$ and $\tilde S^{\rm conf}$ is logarithmic.
More precisely, $S^{\rm conf}=\xi(\tilde S^{\rm conf})$, with $\xi(x)$ of the form
 $\xi(x)=-3N\ln(B-x)+C$, with constants $B$ and $C$.
This suggests that R\'enyi's entropy functional is the right one to start with.
Indeed, let $\alpha=2-q$. The relation between R\'enyi's $I_\alpha(f)$ and $I(f)$, as given by 
(\ref {dual:entropydef}), is $I_\alpha(f)=\xi(I(f))$ with
\be
\xi(x)=-\frac 1{1-q}\ln\left[(2-q)\left(1-(1-q)x\right)\right].
\ee
Take the constant in (\ref {ex:sappr}) equal to $1/(1-q)$. Then one obtains
\be
S^{\rm conf}\equiv
\xi\left(\tilde S^{\rm conf}\right)\simeq\left(\frac {3N}2-1\right)
\left[\ln \frac {U^{\rm conf}}{6N}
+\mbox{ constant}\right].
\ee
This yields
\be
\frac {{\rm d}S^{\rm conf}}{{\rm d}U^{\rm conf}}
=\left(\frac {3N}2-1\right)\frac 1{U^{\rm conf}}=\frac {3N-2}U,
\ee
which is an acceptable relation for the inverse temperature $\beta$.

\section{Discussion}

The main purpose of the present paper is to point out that the
configurational probability distribution of a classical gas
always belongs to the $q$-exponential family. The non-extensivity
parameter $q$ is given by
\be
\frac 1{1-q}=\frac 32N-1,
\label {disc:nep}
\ee
where $N$ is the number of particles. The latter expression has
appeared quite often in the literature, see for instance \cite {PP94,AMPP01,AMAA03}.
The main consequence of our observation is that the subsystem
of configurational degrees of freedom can be described using
standard thermodynamics involving the entropy
$S^{\rm conf}$ of the configurational subsystem.
The thermodynamic relation
\be
k_B\beta=\frac {{\rm d}S^{\rm conf}}{{\rm d}U^{\rm conf}}
\ee
then defines the microcanonical inverse temperature $\beta$.
The same thermodynamic relation can be used to rederive
a known result, which states that the average kinetic energy $U^{\rm kin}$
of a classical gas in the microcanonical ensemble can be obtained
from the density of states $\omega(U)$ and its integral $\Omega(U)$
--- see (\ref {therm:avkinenrg}).

Note that the probability distributions determine the entropy functional
only up to a monotonically increasing function $\xi(x)$.
Therefore we used the notations $\tilde S$ instead of $S$ and $\theta$
instead of $\beta$ for the entropy, respectively the inverse temperature,
because the function $\xi$ which relates $\tilde S$ to the thermodynamic $S$ is
not known {\sl a priori}. However,
in the limit of a large system ($N\rightarrow\infty$) the
nonextensivity parameter goes to 1 and the configurational
probability distribution approximates a Boltzmann-Gibbs
distribution. It is then obvious that $\xi(x)$ should
be chosen in such a way that the inverse temperature $\theta$
becomes the inverse temperature $\beta$ of the canonical ensemble.

For the example of $3N$ non-interacting harmonic oscillators
we calculate all quantities explicitly and obtain an explicit
expression for the scaling function $\xi(x)$. It turns out that
the relevant entropy function, which equals the
thermodynamic entropy when evaluated in equilibrium,
is not that of Tsallis but rather that of R\'enyi.
From the point of view of Jaynes' maximum entropy principle
both are of course equivalent. But in the context of thermodynamics only
one of the two can yield correct values for the temperature.

In conclusion, we did not solve the problem of fixing the right
expression for the microcanonical entropy $S$. But a good
candidate for the configurational contribution $S^{\rm conf}$
is the value of R\'enyi's entropy function evaluated at
the equilibrium value of the configurational probability distribution.

\section*{Acknowledgements}
We are grateful to Prof. Constantino Tsallis for pointing out reference \cite {AMAA03}.


\end{document}